# Leveraging User Profile and Behavior to Design Practical Spreadsheet Controls for the Finance Function

Nancy Wu
www.backofficemechanics.com

**Abstract:** Recognizing that the use of spreadsheets within finance will likely not subside in the near future, this paper discusses a major barrier that is preventing more organizations from adopting enterprise spreadsheet management programs.  But even without a corporate mandated effort to improve spreadsheet controls, finance functions can still take simple yet effective steps to start managing the risk of errors in key spreadsheets by strategically selecting controls that complement existing user practice

**Spreadsheet Usage Continues to Thrive in Finance.**

Spreadsheets are an indispensable tool for organizations today, particularly in finance. Due to their flexibility, computing power and widespread availability, spreadsheets are often the answer for finance functions that need to supplement their financial system(s) quickly and inexpensively.  It is often easier, faster and more cost effective to create an analysis in Microsoft Excel or similar applications than to reconfigure an enterprise resource planning ("ERP") system to perform the same task.  In the last ten years, the heavy reliance on spreadsheets for financial reporting, budgeting, forecasting and planning purposes are consistently observed and echoed by researchers, scholars and business practitioners alike [Panko 1998, Rittweger 2010].  And there is little sign of this trend abating in the immediate future.  In 2011, the Financial Executive Research Institute and Robert Half Management Resources jointly published *Benchmarking the Finance Function* (Thompson 2010).  Of the 207 public and private sector senior executives surveyed for this benchmarking study, more than 60% of finance functions:

- Still relies on Excel for budgeting and planning
- Still relies on Excel for long term planning
- Do not use an account reconciliation tool or system to reconcile accounts.  As manual account reconciliations are mostly completed in Excel, the author deduces that these organizations continue to use spreadsheets to reconcile accounts.

While it might not be possible or desirable to phase out all spreadsheets, organizations can still enjoy the benefits of such a powerful and versatile application by recognizing and mitigating the risks associated with a poorly controlled spreadsheet environment. The intent behind a spreadsheet management program is to treat important spreadsheets as stand-alone applications and apply the same due diligence as other enterprise applications during development and in production.  In other words, verifying that business critical spreadsheets have built-in controls to validate input data, ensure logic accuracy, and manage changes to the spreadsheet's infrastructure.

**While the Risks are Acknowledged, Many Organizations Have Not Implemented Formal Spreadsheet Management Programs.**

While more and more organizations, including several published as EuSpRIG case studies[1] are demonstrating that a spreadsheet management program can be implemented in an efficient and practical manner at both ends of the cost spectrum, many more organizations continue to operate in a loosely controlled spreadsheet environment. Cost of consulting and software licenses aside, their hesitancy to change this status quo may be attributed to the belief that an effort to govern something that impacts so many people on a day-to-day basis cannot help but be intrusive to end users and hence disruptive to the business. Especially among organizations that have not experienced a significant spreadsheet incident, a program that restricts how spreadsheets are created and used may appear to bear too large of an impact upon individuals in return for benefits that, while readily acknowledged (e.g., reduced risk, increased productivity), cannot be easily quantified to build a business case. In addition, spreadsheets belong to no single department, no single function. Rather, they belong to individuals and groups within all functions, all departments. And understandably, these organizations may find that the challenges that come with an initiative to control an application without a central owner are not insignificant. Other spreadsheet risk practitioners echoed similar sentiments in a prior case study: "Ultimately we formed the view that the central difficulty in controlling end user applications is not discovering them or mitigating their risks…The central difficulty lies in formulating, and getting accepted, an organizational response...All aspects of [the end user computing] project required central coordination, but there is as yet no consensus as to the appropriate location of that coordination function. Despite all the evidence of risk and inefficiency, few managers set about addressing these problems, largely, we believe, because they lack the skills and support to do so (Chambers 2008).

**Success of *Informal* Spreadsheet Initiatives Hinges Largely Upon User Acceptance.**

For finance and other functions in want of stronger spreadsheet governance, but find their organizations unable to launch an organization wide program, it is still possible to improve the spreadsheet environment by implementing within the function one specific component of an end-to-end spreadsheet program[2].

One specific component in building an effective program is to *define the set of controls* that will govern spreadsheets. With few exceptions, these controls occur within the spreadsheet themselves. And once defined, the expectation is that they will be built into critical[3] spreadsheets. There are numerous controls that can be effective at preventing and detecting errors when incorporated into spreadsheet design and usage. But upon introduction, certain controls are more likely to be assimilated into normal use (and hence become sustainable) while other controls meet more user resistance. Not surprisingly, spreadsheet controls that fall into the former category are often those that complement existing user practice. In the absence of a corporate edict in the form of a formal spreadsheet program that mandates compliance, minimizing intrusion to the stakeholders through careful spreadsheet control design is critical to gaining user acceptance and making the controls "stick."

**The Finance and Accounting Function's Profile as it Relates to Spreadsheets**

It is the author's experience that many finance and accounting functions do not consistently utilize professional developers to design spreadsheets that process critical financial data. Instead, the task falls upon c*asual spreadsheet designers* – individuals with varying degrees of Excel expertise who lack formal instruction in programming and systems design. Very often they are also the users or reviewers of the spreadsheets who must create spreadsheets to support their primary job functions, such as reconcile balance sheet accounts or review the budget to actual variance at month-end. Some of these individuals have undergone employer sponsored Excel training. While others do not have access to formal training programs through work, the internet provides free content that is easily searchable for anyone who is interested in learning Excel. But often, both formal corporate programs and self-study content (e.g., the internet or "how to" books) primarily emphasize features and functionalities in Excel and do not focus on spreadsheet design from a risk and internal controls perspective[4]. In finance, there also tends to be a preponderance of review and due diligence activities that are done "offline" on paper. Once completed, spreadsheets and supporting documents are printed and submitted for review. The reviewer relies on the printed versions of the spreadsheet to look for mistakes and provide feedback. Because of these user characteristics, the most implementable controls will be those that:

1. Require no advanced Excel technical knowledge – While some casual spreadsheet designers have a strong interest in building Excel expertise, most spreadsheet users and creators just want to "know enough to get by" in order to focus on their core competencies within finance. Introducing controls that require more advanced technical skill does not fit the profile of the typical user and will have a higher likelihood of creating resistance.

2. Enhances review activities on paper – Controls that are deliberately designed to complement and improve review activities that are performed on paper can be placed into use immediately without disrupting existing processes. Consequently, controls that require the individual to change habitual practices will require more education and persuasion.

**Three Examples of Suitable Spreadsheet Controls**

Based on the profile illustrated above, three examples of powerful spreadsheet controls that fit the above criteria are described below. Note that this should not be viewed as an effort to extract the "best" spreadsheet controls in any and all circumstances, but rather it is a demonstration of how understanding the behaviour of a targeted audience can shape the spreadsheet management approach so that internal control recommendations are more practical and relatable to the individuals who must work with them.

*Data Validity Checks for Key Inputs and Calculations (a.k.a. Check Cells)*

The objective of employing data validity checks, or check cells, is to validate accuracy or reasonableness of a body of data. Depending on the level of precision demanded by the user, check cells which come from a separate source are inputted into designated cells and compared to the data from the main spreadsheet body. How comparable the values are to each other (and in some cases, whether the values match exactly) allows the user to evaluate whether the main spreadsheet data produced the expected and correct values, or if an error may have occurred. Examples of check cells include hash totals that should agree with number of input cells imported into the file or a value from an independent source. Additional examples for financial models are described in *Self-Checks in Spreadsheets: a Survey of Current Practice* by David Colver of Operis Group. Designing validity checks to "gut check" portions of the spreadsheet is not only helpful to the user when updating the spreadsheet, but they are especially good for the reviewer. Seeing the results of various sets of cross-checks from secondary sources improve the quality and efficiency of the review process since independently derived conclusions can help the reviewer frame his or her evaluation of the spreadsheet data.

*Clear Data Placement and Labels*

The objective of promoting clear data placement and labels is to treat spreadsheets more as applications (i.e., formal) and less as digital scratch pads (i.e., informal) simply through better data organization. Examples of good organization include grouping all input cells to one area and calculation cells in a separate area. Another example includes visually distinguishing input cells, logic cells and output cells for the reader. Examples of good labeling include clearly demarcating column or row content and unit of measure (currency v. volume, thousands v. millions, etc.). Also, bodies of data that are outdated or no longer relevant should be removed from the spreadsheet. A well-organized and clearly labeled spreadsheet is not only less prone to misunderstanding and errors and more efficient to review on paper, but the file also becomes more transferable as new users can learn the contents of the spreadsheet more quickly than with data that is jumbled and vaguely identified.

*Display of Constants*

First mentioned in the *New Guidelines for Spreadsheets* [Raffensperger, 2000], the objective of displaying constants is to clearly call out the assumptions that feed into the spreadsheet's calculations. Spreadsheets that display constants have key manual inputs such as growth projection rates, interest or discount rate and other assumptions in identified in dedicated and distinct cells. Subsequent calculations made based on these constants then reference the cells using formulas. The opposite of this practice is to place numeric constants *within* the formula cells. The constants become part of the formula string and in effect become "buried." Displaying constants in dedicated and distinct cells increases user efficiency and accuracy as changes in constants need only to be updated once to affect multiple calculations. Reviewers also benefits as they can easily see the assumptions used right on the spreadsheet instead of being hidden within formulas.

**A Case Study**

The three controls illustrated above are not only effective but are also simple and straightforward to implement. So it is not unusual to spot them in certain critical spreadsheets on occasion. But even among finance functions that have already taken informal steps, whether knowingly or otherwise, to build in these controls, tremendous improvement opportunities still exist through more consistent and thoughtful implementation. To illustrate, recently the author was invited by the finance functions of two publicly traded U.S. companies to review a selection of spreadsheets that were considered critical to finance and accounting operations and to provide recommendations on improving the spreadsheet environment. Both organizations were concerned about the health of their business critical finance spreadsheets. Although neither organization had experienced a financial statement error caused by faulty spreadsheets, each group wanted to better understand the control mechanisms within the spreadsheets that would detect any material errors and prevent such an incident from occurring.

While the scope of the review encompassed a broader examination of these spreadsheets' design, high level findings as well as the review criteria pertaining to the three particular controls are summarized below.

*Table 1: Results for Publicly Traded Global Financial Services Organization (2010)*

| Nature of Spreadsheet | File Creation Date | Total No. of Occupied Cells | No. of Calculation Cells | Data Validity Checks for Input and Calculation | Clear Data Placement and Labels | Display of Constants |
|---|---|---|---|---|---|---|
| Cash Flow (1 of 2) | 2008 | 4,146 | 1,427 | **No** | **No** | **Yes** |
| Cash Flow (2 of 2) | 2003 | 2,146 | 1,637 | **No** | **No** | **Yes** |
| Consolidated Financial Statements | 2008 | 21,684 | 41 | **Yes*** | **No** | **Yes** |
| Equity Rollforward | 2005 | 1,924 | 1,334 | **No** | **No** | **Yes** |
| Other Financials Document #1 | 2004 | 1,436 | 1,085 | **No** | **No** | **No** |
| Trial Balance | 2005 | 301,787 | 284,374 | **Yes** | **Yes** | **Yes** |
| Other Financials Document #2 | 2007 | 307 | 38 | **No** | **No** | **No** |
| Financial Statement for 10Q Filing (Select Portion) | 2005 | 1,329 | 283 | **No** | **Yes** | **Yes** |

*Calculation check cells were deemed unnecessary in this case.*

*Table 2: Results for Publicly Traded Global Manufacturing Organization (2011)*

| Nature of Spreadsheet | File Creation Date | Total No. of Occupied Cells | No. of Calculation Cells | Data Validity Checks for Input and Calculation | Clear Data Placement and Labels | Display of Constants |
|---|---|---|---|---|---|---|
| Average Shares Outstanding Calculation | 2000 | 56,500 | 38,500 | **No** | **Yes** | **Yes** |
| Lease Amortization Calculation | 2006 | 420,834 | 122,040 | **Yes** | **Yes** | **Yes** |
| Inventory reserve calculation (1 of 4) | 1997 | 2,523 | 883 | **No** | **No** | **No** |
| Inventory reserve calculation (2 of 4) | 1999 | 2,178 | 1,590 | **Yes** | **No** | **No** |
| Inventory reserve calculation (3 of 4) | 1999 | 106 | 65 | **No** | **No** | **No** |
| Inventory reserve calculation (4 of 4) | 1999 | 1179 | 259 | **No** | **No** | **No** |

*Table 3: Review Criteria*

| Condition | Criteria for "YES" | Criteria for "NO" |
|---|---|---|
| Data Validity Checks for Input and Calculation | Check cells are placed within the spreadsheet to verify the accuracy of inputs and the reasonableness of calculated outputs. In spreadsheets with both inputs and calculations, both types of check cells must be present. | Check cells do not exist or are inadequate. Examples of inadequate check cells include:<br>- Circular referencing<br>- Broken link / Error value<br>- Check cells built in for inputs but not for calculated outputs. |
| Clear Data Placement and Labels | Spreadsheet houses only data required to execute the task. Cell values are clearly labelled and presented logically. Calculation cells are not mixed in with input cells unless they are visually distinct. No hidden data. | Disorganized representation of information. Data is hidden in sheets, columns and/or rows. Existence of irrelevant or out-dated bodies of data. |
| Display of Constants | Hard coded values are displayed independently and labelled in the spreadsheet. Formulas reference the cell location. | Hard coded values are embedded in formula construction without clear labelling. For example: "=780000*.35" where "780000" is the operating income and ".35" is the corporate tax rate. |

**What the Results Mean**

*Table 4: Rate of Compliance with Review Criteria (Derived from Test Results in Tables 1 and 2)*

| Company | Review Criteria | | |
|---|---|---|---|
| | Data Validity Checks for Input and Calculation | Clear Data Placement and Labels | Display of Constants |
| Publicly Traded Global Financial Services Organization | 25% | 25% | 75% |
| Publicly Traded Global Manufacturing Organization | 33% | 33% | 33% |

At the time of fieldwork, neither organization had adopted an end-user computing policy. Yet the results illustrate that each of the three controls was implemented successfully in at least two spreadsheets in both organizations, thus further validating their suitability and ease of implementation for the finance function. The results also show that implementation is inconsistent. In total, only two spreadsheets (13% of total) contained all three controls, and five spreadsheets (36% of total) had none of the three built into the file. So while both groups were already designing these three simple yet powerful controls into key spreadsheets, there are still many opportunities to incorporate them into all critical spreadsheets.

**Conclusion**

Spreadsheet usage continues to thrive in finance with little signs of slowing. In a recent benchmarking study, more than 60% of finance executives surveyed still rely primarily on Excel for budgeting, long term planning and account reconciliations [Thompson 2010]. An effective spreadsheet management program can lead to more reliability in the financial reporting process, less time dedicated to manually intensive mechanical review as well as more capacity available for other value-added activities in finance. But the perceived disruption to the organization, particularly to spreadsheet users and reviewers, and the decentralization of spreadsheet ownership may be preventing more organizations from taking action at an enterprise level.  Absent a formal effort to manage important spreadsheets, finance functions in want of stronger spreadsheet governance should not be discouraged to act.  Because by understanding the function's user profile and tailoring spreadsheet controls to complement user practice, it is possible for finance to successfully implement simple yet effective spreadsheet controls within its own sphere of influence, such as Data Validity Checks, Clear Data Placement & Labels and Display of Constants.  While these measures are straightforward and undemanding in nature, the author's experience shows that they are far from being fully exploited by organizations today.

**Footnotes**

1. *A Practical Approach to Managing Spreadsheet Risk in a Global Business* (T. Lemon, E. Ferguson), *Changing User Attitudes to Reduce Spreadsheet Risk* (D. Balson), *Transforming Critical Spreadsheets into Web Applications at Zurich Financial* (S. Dewhurst), *Spreadsheets - the Good, the Bad and the Downright Ugly* (A. Dunn).
2. An example of an end-to-end spreadsheet management program is described in *A Practical Approach to Managing Spreadsheet Risk in a Global Business* (T. Lemon, E. Ferguson).
3. A spreadsheet's criticality is determined by how important the spreadsheet's purpose and output is to the organization.  Within finance, the most critical spreadsheets are typically those that can materially affect the financial statements.
4. Concluded based on the author's professional experience.  As further evidence, the author inserted "Excel Training" in Google's search bar and reviewed the training content offered by the top three search results and the top sponsored result as listed in the companies' websites: CTS Training, Free Training Tutorial Dot Com, Ozgrid and Microsoft Office – Excel 2007 Training Courses.  All were primarily focused on functionality.